# The Feshbach resonance and nanoscale phase separation in a polaron liquid near the quantum critical point for a polaron Wigner crystal

**M Fratini, N Poccia and A Bianconi**

Department of Physics, University of Rome "La Sapienza", P. le A. Moro 2, 00185 Roma, Italy

E-mail: antonio.bianconi@roma1.infn.it

**Abstract**. The additional long range order parameter that competes with the high $T_c$ superconductivity long range order is identified as an electronic crystal of pseudo Jahn-Teller polarons beyond the critical value of the electron lattice interaction. We show that the region of quantum critical fluctuations in the two variables phase diagram of cuprates: the doping $\delta$ and the chemical pressure (i.e., the tolerance factor, or the average ionic radius of A-site cations) can be measured via the microstrain $\varepsilon$ of the Cu-O length in the $CuO_2$ lattice. The fluctuating order in the proximity of the microstrain quantum critical point that competes with the superconducting long range order is the polaron electronic crystalline phase called a Wigner polaron crystal and the variation of the spin gap energy as a function of microstrain provides a strong experimental support for this proposal.

## 1. Introduction

Quantum phase transitions (QPT) have been identified in different systems going from magnetic materials, heavy fermions [1, 2] and ultra cold gases in optical lattice [3]. The macroscopic phase transition in the ground state of a many-body system occurs when the relative strength of two competing energy terms is varied across a critical value of a coupling term [4]. For a superfluid system at a temperature of absolute zero, by tuning a generic coupling at a critical value $g_c$, a quantum critical point (QCP) appears where the superfluid long range order competes with a second different long range order. At the QCP the thermal fluctuations of standard phase transitions are replaced by quantum fluctuations driven by the Heisenberg uncertainty principle in the ground state.

The search for the mechanism, that allows a quantum macroscopic condensate to resist to the decoherence effects of high temperature, has been focusing on the identification of the critical point of a quantum phase transition [5]. The scientific community has been puzzled by the questions: what is the nature of the long range order that competes with the high $T_c$ superconducting (HTcS) phase? Which is the coupling g that drives the system to the quantum critical point at $g_c$?

There are two main proposals for high $T_c$ superconductivity: the importance of a strong Jahn-Teller electron phonon interaction with the formation of anti Jahn-Teller bipolarons was the Müller driving idea for the discovery of the high $T_c$ superconductivity [6.7], while the importance of electron-electron interaction (the on site Hubbard repulsion U) driving the electronic system to the border of a Mott insulator was stressed by Anderson [8].

Following the Anderson proposal, many authors have addressed their interest to the magnetic spin ordered phases near a Mott insulator and have been looking for the quantum critical point QCP at a critical density of doped holes per Cu site (doping) in the phase diagram where the critical temperature is plotted versus the doping. The superconducting phase has been related with a quantum phase transition to a Mott antiferromagnetic insulating phase [5], with spin fluctuations investigated mainly by inelastic neutron scattering [9], the critical behavior of the spin fluctuations has been measured [10, 11] and other authors have associated the pairing mechanism with spin fluctuations [12].

On the other side following the Muller proposal that the electron lattice interaction is the driving force, the additional long range order (in competition with the HTcS order) has been identified with a commensurate polaron crystal, CPC, that can be described as a generalized Wigner polaron commensurate CDW or a Wigner polaron crystal that shows up at a critical electron-lattice interaction and a critical charge density [13-16].

The classical physics of polarons for the high electron lattice interaction was developed [17-19] for a single polaron or bipolaron (a small polaron at strong electron lattice interaction or a large polaron at weak electron lattice interaction). However at high polaron density the physics of a polaron or bipolaron liquid relevant for HTcS was not clear. Increasing the polaron density several scenarios could show up such as crystal instability, polaron dissociation, phase separation, polaron strings, polaron bubbles, polaron pinning at defects, and the polaron electronic crystalline phases like commensurate or incommensurate CDW and Wigner crystals that could be formed in the presence of long range Coulomb interaction V [20].

## 2. Polarons in cuprates

The local lattice fluctuations due to polarons in cuprates have been determined by the fast and local experimental probes: EXAFS and XANES that probe the instantaneous (with measuring time $10^{-15}$ s) local (in the range of 500 pm) lattice distribution without time or spatial averaging. The solid state scientific community dealing with homogenous crystalline solids was not so familiar with these novel methods developed in the eighties using synchrotron radiation to probe complex inhomogeneous fluctuating systems like glasses, liquids, and biological macromolecules. These methods have been detected the polarons confirming the Muller predictions on the importance of the Jahn-Teller interaction, but experimental results have provided several unexpected results and surprises

1) The undoped cuprate perovskites are correlated charge transfer oxides (where the correlation gap is controlled by the Coulomb repulsion between a hole on copper and another one on the nearest oxygen $U_{dp} \approx 2$ eV) and they are not Mott insulators (where the correlation gap is $U_{dd} \approx 6$ eV between two holes in the same copper ion) ;
2) The dopant holes go into the O(2p) orbital and not into the Cu(3d) orbital as it was expected.
3) The polarons are pseudo Jahn-Teller (pJT) polarons involving the $Q_2$ Jahn-Teller mode called also the half-breathing mode of the oxygen motions that are coupled sterically with the $Q_4/Q_5$ tilts modes and are not of anti Jahn-Teller type.
4) The polarons are in the intermediate coupling regine where the lattice distortion involves a domain of about 8 Cu sites while they were expected to be small polarons in the strong coupling limit localized on a single site.
5) The metallic and superconducting phase shows an unusual nanoscale phase separation and not an homogeneous phase
6) There is coexistence of 1) pJT intermediate polarons condensed in a one dimensional incommensurate charge density wave and 2) itinerant large polarons.
7) The nanoscale phase separation gives an heterogeneous nanoscale material with the nanoscale architecture having a spatial periodicity as large as the superconducting coherence length.
8) A polaron Wigner crystal was observed in a particular perovskite family at hole doping 1/8.

In 1987, XANES experiments have shown that the dopant holes do not form the expected $3d^8$ itinerant states (giving anti-Jahn-Teller polarons) but the dopant holes go in the oxygen 2p orbitals, called ligand holes $\underline{L}$(k) [21], providing a scenario where the O($2p^5$) holes move in a 2D network of metallic oxygen diagonal wires intercalated by the antiferromagnetic 2D spin lattice in of the static holes Cu($3d^9$) in the Cu sites. In 1988-1990 polarons associated with the $3d^9\underline{L}$ singlets have been found to be of pseudo Jahn-Teller (pJT) type [22-27] where the hybridization between the $d_{x^2-y^2}L(b_1)$ and $d_{z^2-r^2}L(a_1)$ orbital is associated with the rhombic distortion of the $CuO_4$ square plane. The pseudo Jahn-Teller polarons in the cuprates are strongly affected by the relevant electronic correlation energies that determine the charge transfer gap in the parent undoped compounds: the interatomic Hubbard repulsion $U_{dl}$ between the dopant hole in the $O(2p^5)$ orbitals and the hole in the Cu($3d^9$) $d_{z^2-r^2}$ or $d_{x^2-y^2}$ orbitals.

In the years 1990-1993 the focus of the research shifted to probe directly the dynamical fast local lattice fluctuations between different Cu site conformations associated with pJT polarons in metallic cuprates focusing on the difference $\Delta R_{apical}$ between the Cu-O(apical) and the Cu-O(planar) bond length that determines the JT energy splitting $\Delta_{JT}$ between $d_{z^2-r^2}$ and $d_{x^2-y^2}$ orbital. In fact the electron-lattice interaction of the pseudo JT polaron type is a complex function of several lattice parameters $\lambda = g(\Delta R_{apical})f(Q)h(\beta)$, where Q is the conformational parameter for the tilts of the $CuO_4$ square planes, and β is the dimpling angle that measures the displacement of the Cu ion from the plane of oxygen ions. The conformational parameter showing the formation of pJT polarons is the splitting $\Delta R_{planar}$ of the in plane Cu-O(planar) bond lengths. The results of EXAFS experiments have provided evidence for nanoscale inhomogeneity with the segregation of stripes of localized pseudo Jahn-Teller polarons and stripes of itinerant carriers [28-32, 13-16]. These unexpected results have been confirmed by further experimental investigations such as thermopower [33-35], high resolution EXAFS [36-43], resonant x-ray diffraction [44] the electron paramagnetic resonance (EPR) of $Mn^{2+}$ doped cuprates [45], copper NQR spectra, demonstrating the existence of a second anomalous copper site in lanthanum cuprate whose character is independent of the method of doping, and systematic NMR/NQR experiments [46, 47], susceptibility measurements [48], local structure investigation using the atomic pair distribution function (PDF) analysis of neutron powder-diffraction [49-50], isotope effects on T* [50-52] and an ultrafast real-space probe of atomic displacements (with sub-picometre resolution), the MeV helium ion channelling, to probe incoherent lattice fluctuations [53], and recently by the anomalous energy distribution curves E(k) (that strongly deviate from band structure calculations) in ARPES and the Fermi arcs measured by k scanning ARPES. [53-66]. There is therefore now a compelling evidence for polarons and nanoscale phase separation in cuprates that make these materials a realization of a particular pseudo Jahn-Teller polaron fluid showing a nanoscale phase separation and local lattice fluctuations [67-71] typical of complex systems and biological systems [72-73].

In 1995 after the compelling evidence for polarons in cuprates was found by fast and local probes the scientific community started to investigate the physics of fluids made of Jahn-Teller polarons focusing on doped manganites that show colossal magnetoresistance [74-85]. It was recognized clearly that these systems show a nanoscale phase separation as in cuprates [35, 86, 87, 88]. The local lattice fluctuations are similar but larger than in cuprates [81]. The energy dispersion and pseudogap in ARPES [64] are similar to the experimental features of ARPES of cuprates and finally in particular materials and at particular dopings x=0.5 electronic crystalline phases called Wigner polaron crystal [82] or commensurate polaron charge density waves [89] show up as it was shown before in particular cuprate families [14-16] and the theory of polaron Wigner crystals is now rapidly developing [90-93].

In the superconducting materials the experimental results show a nanoscale phase separation with the coexistence of incommensurate stripes with fluctuating pseudo Jahn-Teller polarons and superconductivity that have been recently confirmed [94-96].

### 3. Feshbach resonances

For a metal confined in a single membrane Blatt proposed in 1963 [97, 98] that the critical temperature is amplified by tuning the chemical potential at an electronic critical point of the electronic structure (where the Fermi surface of a new subband given by quantum size effects appears or disappears) by a shape resonance in the interband BCS coupling that is similar to the Feshbach resonances in nuclear physics, Recently the shape resonance predicted by Blatt has been observed in a thin Pb films [99-101].

The investigation of Bi2212 superconductor has given provided evidence in 1993 that the high $T_c$ superconductivity occurs in heterostructure at atomic limit where a superconducting material is intercalated by a different material with different electronic structure forming a superlattice of quantum stripes. It was reported that in cuprates a shape resonance (called also Feshbach resonance) where the dimensionality of Fermi surface of one subband changes from 2D to 1D that is a critical point of the electronic structure, driven by the charge density and the architecture of the superlattice. At this critical point a large amplification of the critical temperature is realized driven by the role of the exchange like interband pairing term [102-113]. The Feshbach resonance has been now shown to be the driving mechanism for high $T_c$ in magnesium diboride [114] that is made of a superlattice of graphene-like boron monolayers intracalated by magnesium of aluminum.

In 1993 the Feshbach resonance in ultra cold gases was proposed independently by Teisinga et al, [115] for increasing the bose condensation temperature. Later it was also proposed to realize Feshbach resonances in optical lattices that has interesting similarities with the Feshbach resonances in metallic quantum superlattices [116].

The recent theoretical finding that the Feshbach resonance occurs near the quantum critical point in. atomic Bose gases [117] supports the similarity with the exchange pairing at a shape resonance in metallic superlattices of quantum wires proposed in cuprates [103-114]. In fact there is an increasing number of experiments supporting the idea that in the copper oxide layers of cuprates there is a quantum phase transition from a polaron liquid to an electronic crystalline phase of ordered polarons that was identified as a polaron Wigner crystal by increasing the electron lattice interaction at a critical value $\lambda(\varepsilon_c)$ and at constant charge density $\delta_c=1/8$. Therefore at constant doping increasing the electron lattice polaronic interaction we reach a critical point where a quantum phase transition occurs [13] and the Feshbach resonance occurs in the regime dominated by quantum local lattice fluctuations [118, 119, 120].

### 4. The chemical pressure in new phase diagram

The cuprate perovskites are heterogeneous materials formed by three different building blocks [BO](AO) $CuO_2$: the metallic bcc $CuO_2$ layers, the insulating AO layers (rock-salt fcc layers in hole doped cuprates) (A=Ba, Sr, La, Nd, Ca, Y etc.) and the charge reservoir BO layers. In the perovskite materials like $La_2CuO_4$ (or $[LaO]_2CuO_2$) and manganites $AMnO_3$ (or $[AO]MnO_2$) the lattice matching between the transition metal oxide layer MO (M=Cu or Mn) and that of the rare earth (A=La, Y, Ba, Sr) oxide (AO) is realized by rotating the crystalline axis by $45^0$ and a good matching occurs if the ratio between the interatomic distances $r_{A-O}$ and $r_{M-O}$ (the respective bond lengths in homogeneous isolated parent materials AO and $MO_2$) is $\frac{r_{A-O}}{r_{M-O}\sqrt{2}}=1$ The chemical pressure (or internal pressure) due to the interlayer lattice mismatch across the block-layer interface is usually measured by

$$\eta = 1 - t = \frac{r_{M-O} - (<r_A> + r_O)/\sqrt{2}}{r_{M-O}}.$$

Where the Goldschmidt tolerance factor is $t = \frac{r_{A-O}}{\sqrt{2}r_{M-O}}$. In doped manganites with a simple perovskite $A_{1-x}A'_xMnO_3$ structure the tolerance factor is usually calculated using the average A cation size $<r_A>$ [76] and keeping fixed the ionic radii of oxygen and copper to get $r_{M-O}$.

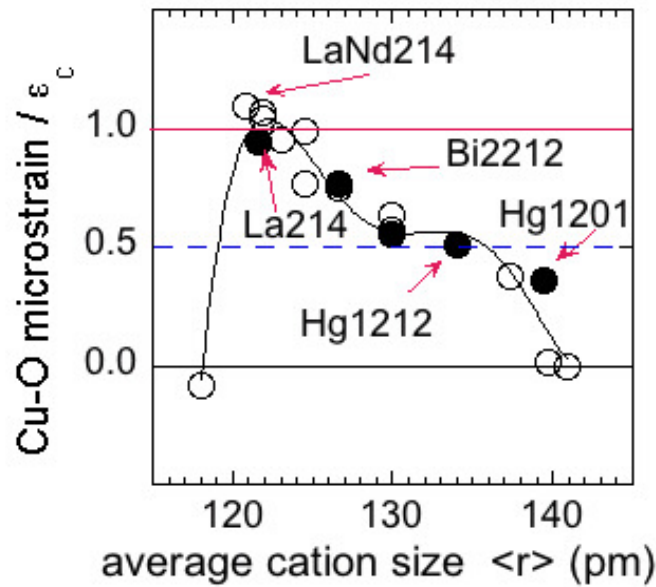

**Figure 1.** The Cu-O micro-strain ε=((197-CuO(pm))/197)) normalized to its critical value (4%) as a function of the average ion size in the rock-salt layers that are first neighbours of the $CuO_2$ plane. The static commensurate pJT polaron crystalline phase in the LTT phase occurs for $\varepsilon/\varepsilon_c > 1$.

There is a large experimental agreement that the magnetic and electronic phases of the polaron fluid in manganites depend not only on the density of the polarons but also on the chemical pressure [76, 86, 121].

In hole doped cuprate superconductors there is agreement that the stress exerted on the $CuO_2$ plane by the lattice mismatch induces a compressive $CuO_2$ microstrain of the Cu-O distance and a tensile microstrain in the rocksalt layer. It is well known that the cuprate perovskite lattice is stable only up to a critical value of the lattice tolerance factor between the $CuO_2$ layer and the rocksalt layers in the $La_{1-x}A_xCuO_2$ family and when the average ionic radius in the rocksalt layers is smaller than 118 pm the system go into the T' phase where the intercalated layers have a fluorite structure [122-127].

The increasing lattice mismatch induces rotation, tilting and dimpling of the $CuO_4$ square planes i.e., corrugations of the flat $CuO_2$ plane. The ordering of these tilts induced by lattice mismatch gives the transitions from tetragonal (HTT) to orthorhombic (LTO) and finally to the LTT phase in cuprate superconductors.

Studies of the effects of the tolerance factor within a single cuprate family have been reported [122-127], mainly for the La124 family, but it was not possible to extend the calculation of the tolerance factor to other complex perovskite families of cuprates (like Y123, Bi2212 Hg1212 and others) where there are multiple different ions and multiple intercalated different layers.

To overcome these difficulties we have proposed in 1998 at the second stripes conference [128] to measure the chemical pressure due to the interlayer mismatch by measuring the average copper-oxygen bond length $<Cu-O>$ by EXAFS or by refinement of the XRD data avoiding the use of the average cation size $<r_A>$, taken from the Shannon tables.

$$\eta = 1 - t \propto k \frac{r_{M-O} - <Cu-O>}{r_{M-O}}.$$

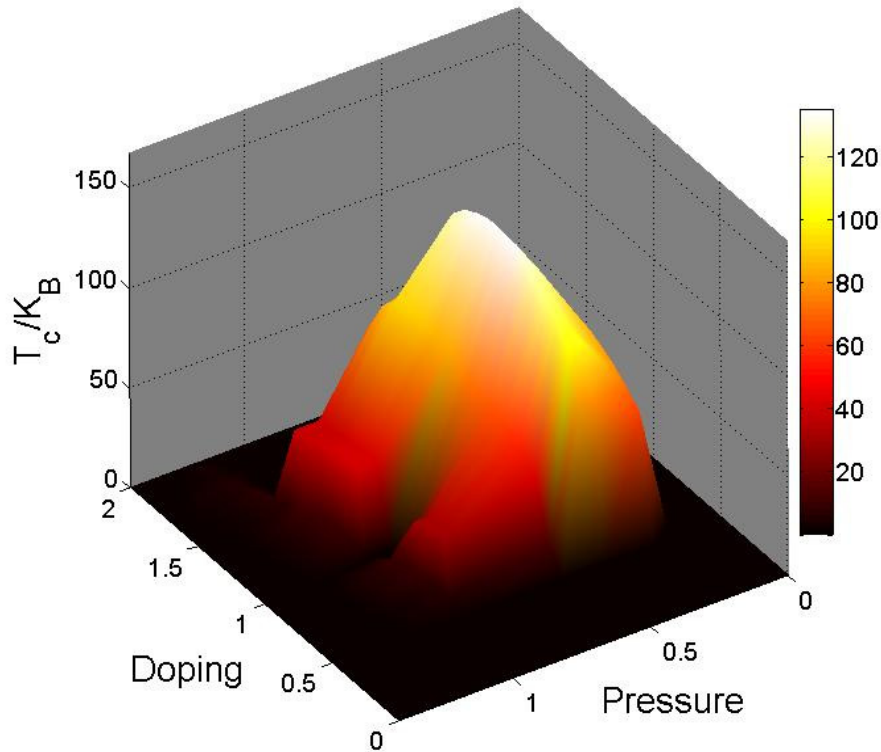

**Figure 2.** The critical superconducting temperature function of doping $\delta/\delta_c$ and of pressure via microstrain $\varepsilon/\varepsilon_c$ normalized at the critical doping ($\delta_c=1/8$) and the critical Cu.O microstrain ($\varepsilon_c=4\%$) for the formation of the commensurate polaron crystal (CPC). The superconducting long range order parameter clearly competes with the commensurate pJT polaron crystal formed at (1, 1) that has been described as a paired Wignet polaron crystal.

where the pressure is measurd via the microstrain in the $CuO_2$ lattice $\varepsilon = (197 - \langle Cu-O \rangle)/197$ and the <Cu-O> bond length is measured in picometers thatt is plotted in Fig. 1.

The idea is based on the use of the unrelaxed equilibrium value for the Cu-O bond length $r_{Cu-O} = 197$ $pm$ measured for the planar $CuO_4$ units made by the free $Cu^{2+}$ ions in water measured by EXAFS. The constant k is determined by the difference of the elastic constants between the copper oxide plane and the intercalated planes (it was taken to be about 2 in previous works). There is a critical value of the microstrain $\varepsilon_c = 4 \pm 0.3\%$ (i.e., a critical average bond length $\langle Cu-O \rangle_c = 189 \pm 0.5$ $pm$). For the values of the microstrain larger than the critical microstrain the crystallographic LTT phase appears and the commensurate static spin modulation appears at doping $\delta_c=1/8$ in neutron scattering experiments [9], The critical microstrain is controlled within the range 3.7-4.3%.by the lattice disorder induced by the variance of the ionic radii of intercalated ions [129].

The lattice interlayer mismatch or chemical pressure measured by the Cu-O microstrain is related with the average ion size in the rocksalt layers in contact with the $CuO_2$ plane. In Fig. 1 the values of the microstrain normalized to the critical value $\varepsilon_c=4\%$ as a function of the average ion size of the cations in the nearest rockalt intercalated monolayers is plotted.

The superconducting critical temperature $T_c(\delta,\varepsilon)$ of many cuprate perovskites as a function of normalized doping $\delta/\delta_c$ and the normalized $CuO_2$ microstrain $\varepsilon/\varepsilon_c$ is plotted in Fig. 2 at ambient pressure. The maximum $T_c$ of 130 K occurs in the mercury cuprate family at doping $\delta/\delta_c = 1.3$ ($\delta=16\%$) and microstrain $\varepsilon/\varepsilon_c = 0.55$ (i.e. at $\varepsilon = 2.2\,\%$ or at the Cu-O bond length of 192.65 pm). The cuprates $La_{0.6}Nd_{0.4}La_{1-x}Sr_xCuO_4$ with microstrain $\varepsilon/\varepsilon_c > 1$ are in the region where Wigner polaron crystal suppresses the superconducting critical temperature around the quantum critical point at (1,1) as it can be seem clearly in the figure. Therefore the electron-phonon coupling, controlled by the chemical pressure, is the variable that drives the system to localization and there is a quantum critical point where an electronic solid with long range order competes with the superconducting order as was required by several theories [130].

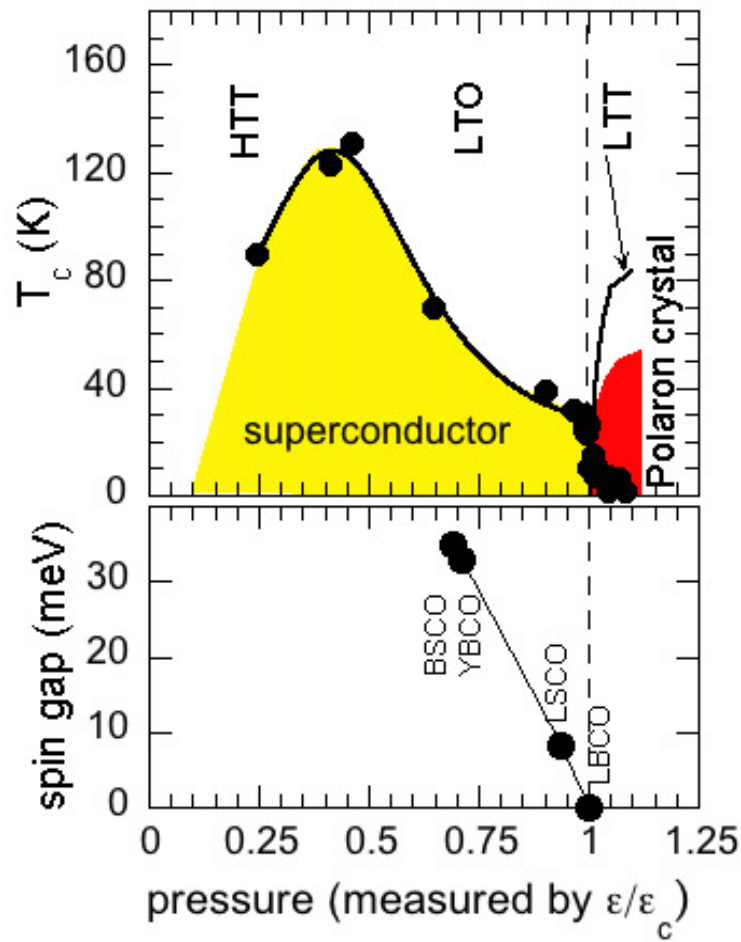

**Figure 3.** *Lower panel:* The spin-gap energy as a function of normalized microstrain. In different superconducting cuprates at optimum doping from ref.[133, 137,138, 139]. *Upper panel*: the superconducting critical temperature as a function of the microstrain at constant doping (1/8) and the Wigner polaron crystal in Nd doped LCO cuprate perovkites from Ref. 13.

The formation of a polaron Wigner crystal [90-93] was first proposed in 1993 [31, 14-16] for a critical polaron density $\delta_c$ and a critical electron-phonon interaction $g_c$ to explain the drop of $T_c$ in $La_{0.6}Nd_{0.4}La_{1-x}Sr_xCuO_4$. and $La_{1.875}Ba_{0.125}CuO_4$. Later it was found that these systems show static spin stripe order [131, 132, 133] characterized by four magnetic diffraction spots at $(\pi \pm \delta, \pi)$ and $(\pi, \pi \pm \delta)$ where δ is close to 0.125 that have been called the "static stripe phase" that we assign to a striped polaron Wigner crystal in fact the magnetic order is related with the lattice structure[134-135]. Starting from $La_{1.875}Sr_{0.125}CuO_4$ where we have a polaron liquid and spin fluctuations the Wigner crystallization can be induced by increasing the microstrain (i.e. decreasing the average ion size) in the case of $La_{1.875-x}Nd_xSr_{0.125}CuO_4$ while in the case of $La_{1.875}Ba_{0.125}CuO_4$ the system close to the critical point is pushed to localization by decreasing the critical strain value by increasing the disorder via increasing the ion size variance The polaron Wigner crystals have been observed also in the nichelate $La_2NiO_{4.25}$ [136] and in manganites $LaSr_{0.5}Ca_{0.5}MnO_3$ [82] at a critical value ($\varepsilon_c$, $\delta_c$).

In order to understand the relevance of the quantum critical point for the formation of the polaron Wigner crystal on the superconducting phase we have plotted in Fig. 3 the energy of the spin gap for the onset of spin fluctuations in the metallic phase of the polaron fluid for values of the microstrain smaller than the critical value. The cuprates show an universal distribution of magnetic fluctuations. A spin gap is observed at low energy in different cuprate families.

Fig. 3 shows the spin-gap energy for several different cuprates near optimal doping. $La_{2-x}Sr_xCuO_4$ (x=0.16) (LSCO) from reference [137], for $YBa_2Cu_3O_{6.85}$ (YBCO) from reference [138]; $Bi_2Sr_2CaCu_2O_{8+y}$ BSCO (estimated by scaling with respect to $YBa_2Cu_3O_{6+x}$ from reference [139] and finally the zero gap for $La_{2-x}Ba_xCuO_4$ (LBCO) from reference [133]. The experimental results show a correlation between magnetic excitations and microstrain that applies to a variety of cuprates. This trend makes clear that the magnetic excitations show the typical behavior near a critical point of a quantum phase transition where the high $T_c$ superconductivity occurs in the region of a quantum paramagnetism near the onset of quantum fluctuations.

In conclusion we have shown that the recent experimental results support the scenario proposed in ref. [13] that the high $T_c$ superconductivity occur near a Quantum Phase Transition for a polaron fluid where the superconducting long range order competes with the long range order of polaron Wigner crystal.

**Acknowledgments**

We acknowledge financial support from European STREP project 517039 "Controlling Mesoscopic Phase Separation" (COMEPHS) (2005).